\documentclass[a4paper,11pt]{article}

\usepackage{pos}
\usepackage{lipsum} %package to generate placeholder text in the following
\usepackage[separate-uncertainty=true]{siunitx}
\usepackage{amsmath}
\title{Updated directions of IceCube HESE events with the latest ice model using DirectFit}

\ShortTitle{12 years of high-energy starting events in IceCube}

\author{The IceCube Collaboration \\{\normalsize \normalfont(a complete list of authors can be found at the end of the proceedings)}\\}

\emailAdd{tyuan@icecube.wisc.edu}
\emailAdd{dima@icecube.wisc.edu}

\abstract{

The initial evidence of astrophysical neutrinos by the IceCube Neutrino Observatory stemmed from the high-energy starting events (HESE) sample: a selection of the highest-energy neutrino interactions that occurred within the detector fiducial volume. Each event was reconstructed based on our best knowledge of the ice at the time, with the latest results published in a description of the sample using 7.5 years of data. Since then, several major improvements in ice modeling have occurred using in-situ calibration data. These include a microscopic description of ice anisotropy arising from ice crystal birefringence and a more complete mapping of the ice layer undulations across the detector. The improvements feed into more accurate descriptions of individual events, and can especially affect the directional reconstruction of particle showers. Here, we apply the latest ice model in an exact manner to reconstruct IceCube’s high-energy events using DirectFit. This reconstruction samples posterior distributions across parameters of interest by performing full event resimulation and photon propagation at each step. We obtain improved per-event descriptions, as well as updates on previously published source searches using the aggregated sample.
\vspace{4mm}
{\bfseries Corresponding authors:}
Tianlu Yuan$^{1*}$, Dmitry Chirkin$^{1}$\\
{$^{1}$ \itshape Dept. of Physics and Wisconsin IceCube Particle Astrophysics Center, University of Wisconsin{\textendash}Madison}\\[4mm]
$^*$ Presenter

\ConferenceLogo{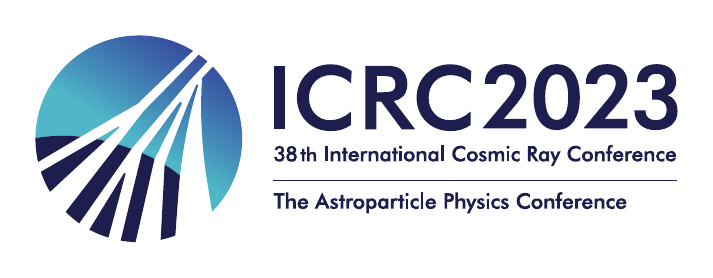}

\FullConference{The 38th International Cosmic Ray Conference (ICRC2023)\\ 26 July -- 3 August, 2023\\ Nagoya, Japan}
}

\newcommand{\like}{\mathcal{L}}

\newcommand{\pdfSxi}{S(\vec{x}_i, \vec{\psi}_s)}
\newcommand{\TS}{\mathrm{TS}}

\begin{document}

\maketitle

\section{Introduction}\label{sec:intro}

Over a decade ago the IceCube Neutrino Observatory detected the first PeV neutrinos~\cite{IceCube:2013cdw}, a crucial piece in the first evidence of high-energy extraterrestrial neutrinos~\cite{IceCube:2013low}. The analysis was performed with the high-energy starting events (HESE) selection, which utilized a simple outer-layer veto coupled with a lower bound on the total detected charge of 6000 photoelectrons in order to suppress atmospheric backgrounds and select high-energy neutrino interactions within a fiducial volume. In the time since, IceCube has continued to collect data nearly without interruption. Iterative updates to the HESE sample were performed as events accumulated~\cite{Kopper:2015vzf,Kopper:2017zzm}. As understanding of the ice and detector evolved, the HESE sample was reanalyzed with an increased sample size and improved models and methods~\cite{IceCube:2020wum}. Included were per-event reconstructed quantities, which for the spatial clustering analyses relied on a fully resimulated event reconstruction~\cite{Chirkin:2013avz} (DirectFit\footnote{Code available at https://github.com/icecube/ppc/}) using an ice model that was current at the time.

Here, we provide an updated neutrino source search using the HESE sample with an extra 4.5 years of data since~\cite{IceCube:2020wum}. In addition to the increased livetime, major updates to modeling of ice anisotropy~\cite{tc-2022-174} and ice layer undulations~\cite{Chirkin:2023icrc} are taken into account. Per-event directional probability density functions (PDF) are again parameterized with the 8-parameter Fisher-Bingham (FB8) distribution to capture generalized, asymmetric PDFs on the sphere~\cite{kent1982,Yuan:2020}. The parameterized description of all events and pre-computed full-sky FITS files of their PDFs are available at~\cite{DVN/PZNO2T_2023}.

\section{Data sample and event reconstruction}\label{sec:reco}

In total, 64 new events passed the HESE selection. Events 128 and 132 contain coincident atmospheric backgrounds and are removed from the analysis. The remainder is combined with the 102 previously reported events~\cite{IceCube:2020wum} for a total of 164 events.

\begin{figure*}[hbt]
\centering
\includegraphics[width=0.49\textwidth,trim={2cm 2cm 1.45cm 2cm},clip]{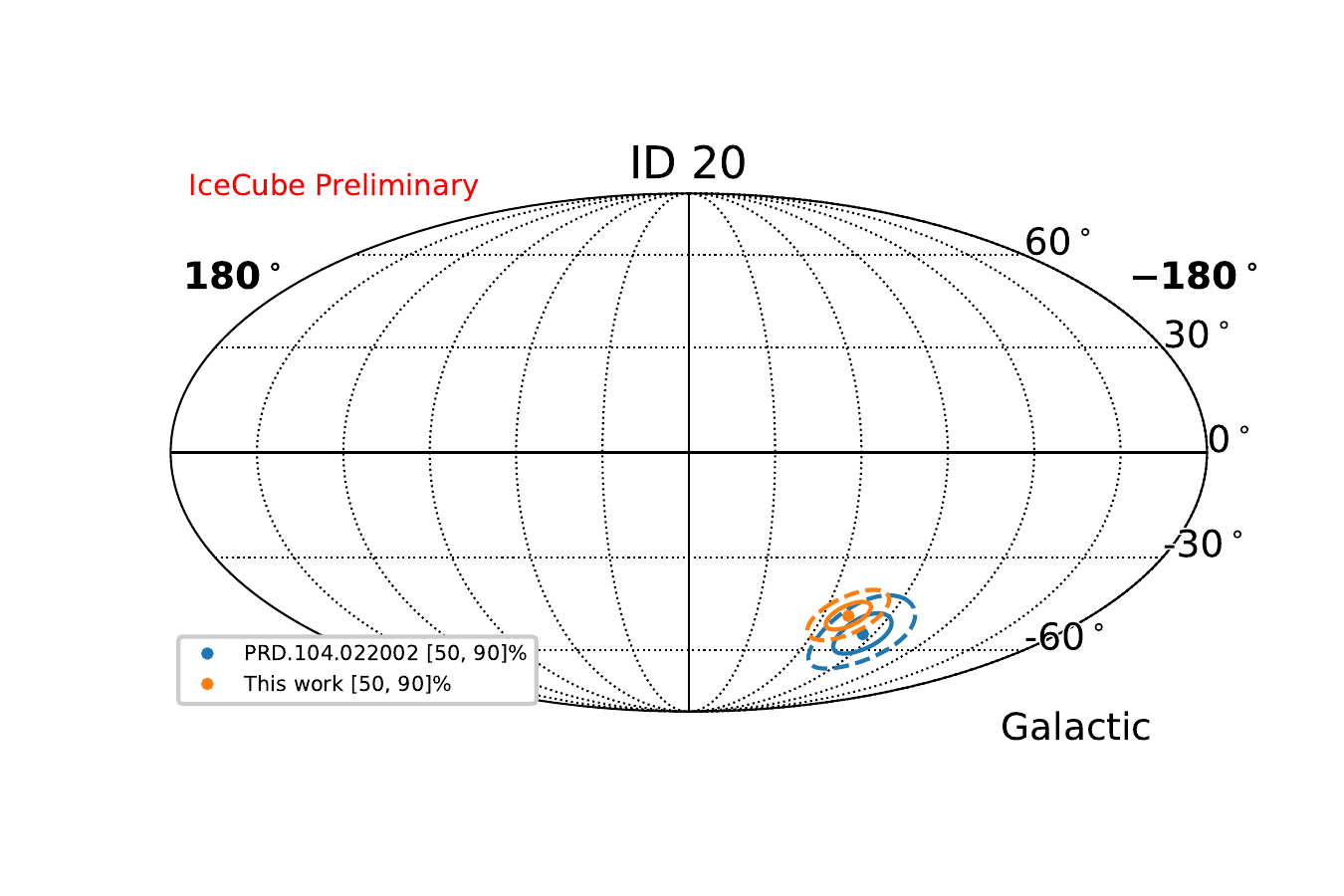}
\includegraphics[width=0.49\textwidth,trim={2cm 2cm 1.45cm 2cm},clip]{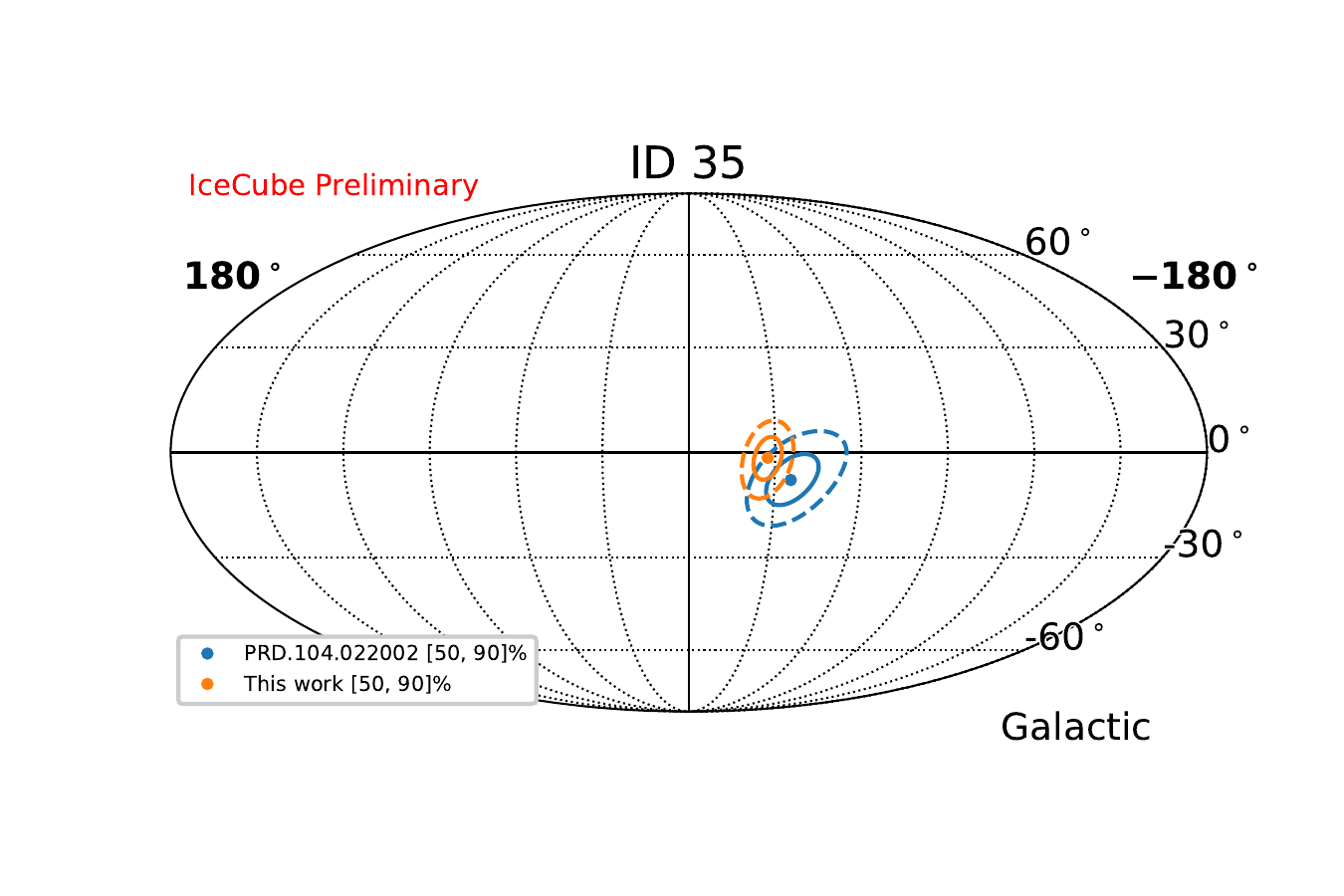}
\caption{Comparison of the reconstructed directions and 50\% (solid) and 90\% (dashed) highest posterior density regions for events 20 (left panel) and 35 (right panel).}
\label{fig:vs}
\end{figure*}
A primary goal of this proceeding is to provide per-event reconstructed quantities, including the direction and its uncertainties as well as the visible energy, using the most up-to-date ice model. This is currently only possible with DirectFit, which resimulates events as it searches through the physics parameter space and fully propagates photons on graphics processing units (GPU) through an ice model in all its descriptive detail. Its accuracy comes at a significant cost of computational complexity; for the energy ranges of HESE events and the settings used here a complete reconstruction can take on the order of hours to a day on a NVIDIA A100 GPU. Reconstructing large scale Monte Carlo (MC) datasets with DirectFit is currently not feasible. Therefore, as remarked in~\cite{IceCube:2020wum}, the results presented here are not meant to be employed in a fully consistent, forward-folding likelihood fit for MC-based spectral analyses. Alternative IceCube reconstructions that apply approximations are generally faster and more applicable to spectral fits~\cite{Basu:2023icrc,Silva:2023icrc,Naab:2023icrc}.

DirectFit was run for all 164 events, first with a localized random search routine to refine the physics parameter space over the event position and its arrival direction. This is followed by a second Approximate Bayesian Computation (ABC) routine to sample the posterior density~\cite{Chirkin:2013avz}. Finally, the distribution of samples is marginalized over position and energy, and a FB8 PDF is fitted to the sample of arrival directions. In comparison to~\cite{IceCube:2020wum}, the reconstruction is updated to have increased simulation photon statistics used in DirectFit as well as the ice model updates discussed earlier.

For the events already present in the previous 7.5-year dataset the reconstructions provided here may be compared to those in~\cite{IceCube:2020wum} (Appendix J). Figure~\ref{fig:vs} shows the 50\% and 90\% highest posterior density (HPD) regions for two PeV events that were previously published (blue)~\cite{IceCube:2020wum} to those obtained in this work (orange). Differences arise from the updates to the ice optical modeling and the smaller contours result in part due to the increased simulation photon statistics.

\begin{figure*}[hbt]
\centering
\includegraphics[width=0.7\textwidth]{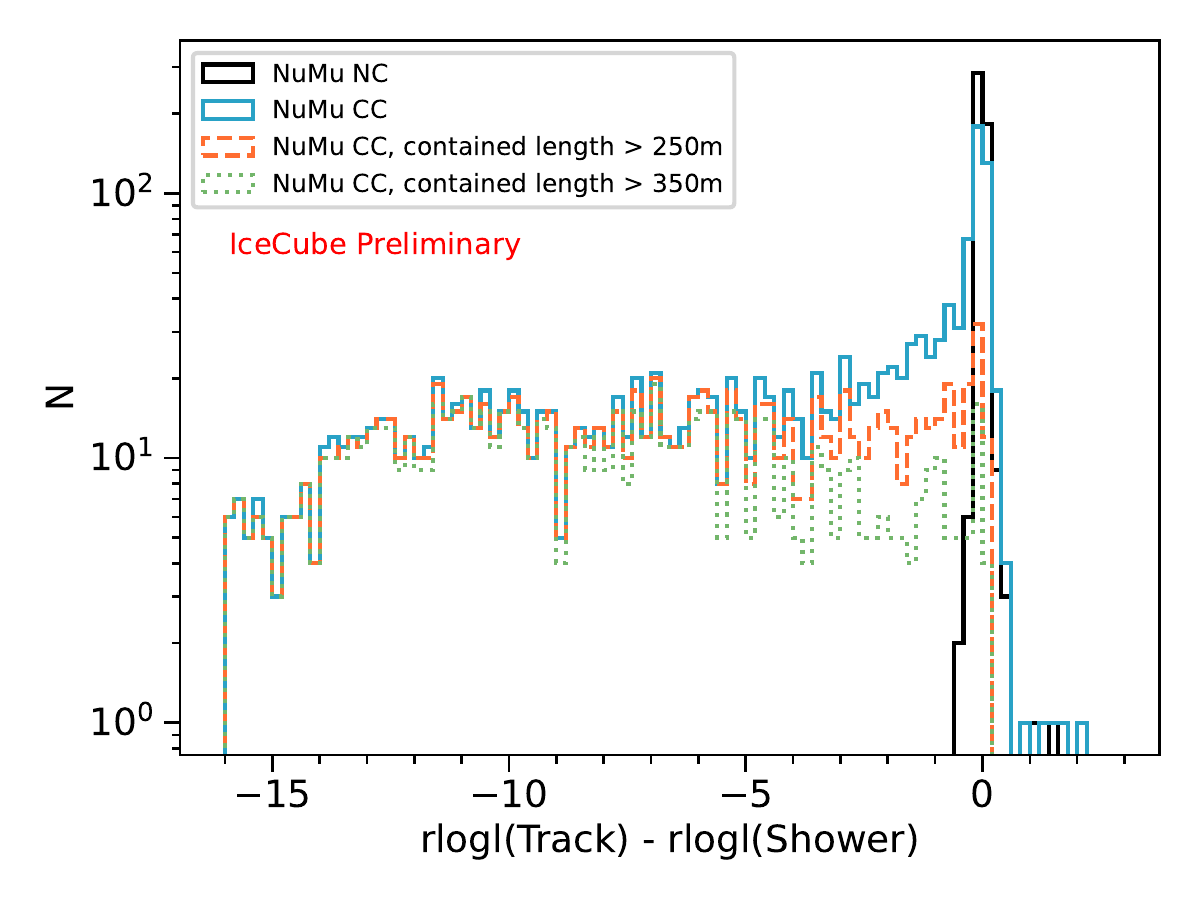}
\caption{Distribution of the reduced negative log-likelihood difference between a track and shower reconstruction for a sample of simulated muon neutrino events. Neutral current (NC) interactions (black) produce hadronic showers and an outgoing neutrino, while charged current (CC) interactions (blue) produce an outgoing charged muon. Separability improves as the contained length increases to greater than \SI{250}{\m} (dashed orange) and \SI{350}{\m} (dotted green).}
\label{fig:dnumu}
\end{figure*}
As DirectFit provides distinct algorithms for reconstruction of particle showers and high-energy muon tracks, events are first classified for the appropriate algorithm. To predetermine which reconstruction to apply, a new classification scheme was implemented using a faster, but more approximate, reconstruction that includes recent ice model improvements~\cite{Aartsen:2013vja,YUAN2023168440,Yuan:2023icrc}. For each event, the fast track and shower reconstruction was run to obtain the reduced negative log-likelihood (rlogl) under each hypothesis. Their difference gives the classification metric, as shown in Fig.~\ref{fig:dnumu} for a simulated sample of muon neutrinos. Interactions via deep inelastic scattering result in a hadronic shower and either a neutrino or a muon in the case of a neutral current (NC) or charged current (CC) interaction, respectively. More negative values correspond to more evident track signatures, and separability against showers improves for tracks with longer contained lengths inside the instrumented volume (dashed orange and dotted green). For HESE data events with $-0.8 < \mathrm{rlogl(Track)} - \mathrm{rlogl(Shower)} < -0.2$, a full DirectFit is then run with both track and shower reconstructions and the one yielding the best data agreement is chosen. This two step approach is taken as it is too computationally expensive to reconstruct every event with DirectFit under both shower and track hypotheses. For many events, such as in the case of an obvious track, running both routines is also unnecessary.

Event 135 is the highest-energy event in the sample. Reconstructed as a track, its most-probable visible energy is \SI{4.8}{\peta \eV}. Assuming that the track is a high-energy muon, we find $dE/dx=\SI{1.125}{\tera \eV \per \m}$ in the last \SI{400}{\m} before exiting the detector. Sampling from a uniform initial muon energy distribution between \SIrange{1}{100}{\peta \eV}, and conditioning on an energy loss of $\SI{450}{\tera \eV}$ after \SI{400}{\m}, the initial muon energy is found to be $\SI{9\pm5}{\peta \eV}$~\cite{Chirkin:2004hz}. This corresponds to a total energy of about \SI{13\pm5}{\peta \eV}. If the event occurred from the CC interaction of a muon neutrino, the incoming neutrino energy would be further dependent on flux assumptions and the inelasticity. Further detailed studies are ongoing.

\section{Spatial clustering analysis}\label{sec:spatial}

We performed three tests of spatial clustering, following a similar procedure as previously reported~\cite{Kopper:2015vzf,Kopper:2017zzm,IceCube:2020wum}. An unbinned, maximum likelihood function is used~\cite{Braun:2008bg}
\begin{align}
	\like(n_s;\vec{\psi}_s) &= \prod_i^N
	\left[
	\frac{n_s}{N}\cdot \pdfSxi
	+ \left( 1 - \frac{n_s}{N} \right)\cdot B (\delta)
	\right],
	\label{eq:pslikelihood}
\end{align}
where $N$ is the total number of observed events, $n_s$ is the expected number of signal events, $\vec{\psi}_s$ is the source position in the sky, $\vec{x}_i$ represents the properties of event $i$, 
 $B(\delta)$ is the declination-dependent background spatial distribution derived from data scrambles along a declination band $\delta$, and $\pdfSxi=P_i(\vec{\psi}_s)$ is the spatial distribution expected from the signal population which is taken to be equal to the posterior density of the event direction at the source position. For spatially extended sources, the likelihood is convolved with the spatial template over the entire sky. The TS is defined to be,
\begin{equation}
\TS=-2\ln[\like(n_s=0)/\like(\hat{n}_s)],
\end{equation}
where $\hat{n}_s$ is best-fit value that maximizes the likelihood. The TS distribution under the null hypothesis is obtained by scrambling data in right ascension to generate pseudoexperiments expected under an isotropic flux.

\subsection{Data-driven sensitivities to the galactic plane}\label{sec:sensitivity}

\begin{figure*}[hbt]
\centering
\includegraphics[width=0.7\textwidth]{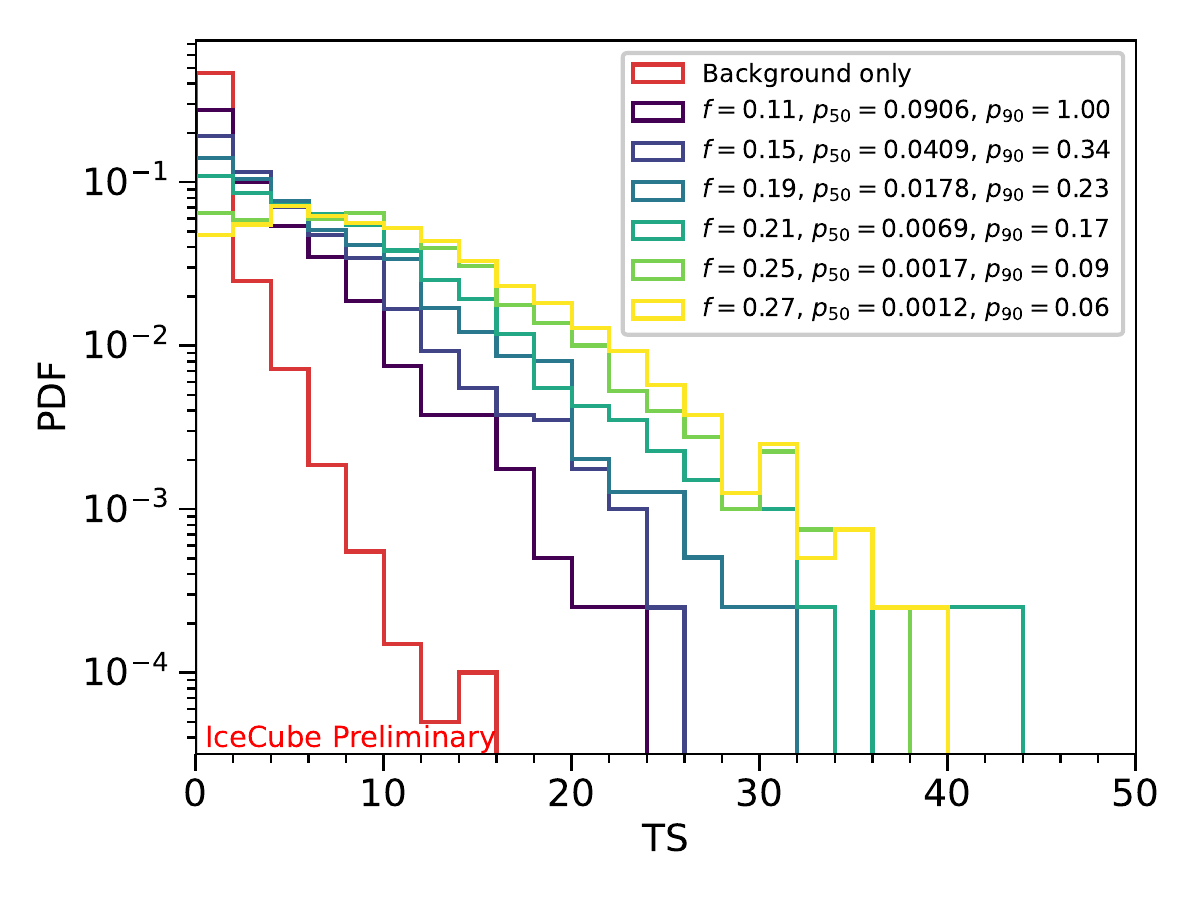}
\caption{The TS distribution obtained for the Fermi-$\pi^0$ galactic plane template for an ensemble of pseudoexperiments at different values of $f$ obtained following the procedure detailed in the text. As $f$ increases TS values tend to increase as more events cluster around the galactic template. The p-values shown in the legend are obtained by comparing the 50\% and 90\% levels (cumulative from the right) against a background TS PDF ($f=0$) and correspond to the probability a higher TS value would be obtained under the null.}
\label{fig:pi0ts}
\end{figure*}
Without MC, a data-driven approach was taken to estimate sensitivity to the galactic plane. The idea is to evaluate the impact of a galactic contribution as a fraction, $f$, of the total diffuse neutrino flux. We assume the best-fit values obtained in~\cite{IceCube:2020wum} for a single-power-law model of the diffuse astrophysical neutrino flux, and the Fermi-$\pi^0$ galactic plane spatial distribution~\cite{Ackermann_2012,Gaggero_2015}. With these assumptions, each of the 164 events in data can be probabilistically categorized as atmospheric, astrophysical diffuse, or galactic thus allowing for generation of pseudoexperiments. 

The procedure is as follows. The atmospheric probability is dependent on reconstructed energy and zenith using the background distributions from~\cite{IceCube:2020wum}. If an event is not categorized as atmospheric, the probability of it being of galactic origin is determined by $f$. For events that are categorized as galactic, a random unit vector, $\mathbf{\hat{u}}$, is sampled from its directional PDF and a second unit vector, $\mathbf{\hat{v}}$, is sampled from the Fermi-$\pi^0$ skymap within a $\pm 10^\circ$ declination band of $\mathbf{\hat{u}}$. The entire PDF is then rotated such that $\mathbf{\hat{u}}$ aligns with $\mathbf{\hat{v}}$, thus simulating that event as being of galactic origin. The entire procedure is repeated 2000 times for values of $f$ ranging from 0.01 to 0.9, and for each ensemble of pseudoexperiments that correspond to a particular value of $f$ a test statistic (TS) distribution can be obtained and compared to that assuming an isotropic flux (Fig.~\ref{fig:pi0ts}). Implicit are assumptions that the reconstructed distributions from~\cite{IceCube:2020wum} approximate DirectFit reconstructed quantities, and that the galactic flux follows the same spectral shape as the diffuse astrophysical flux.

\subsection{Results}\label{sec:results}
\begin{figure*}[hbt]
\centering
\includegraphics[width=\textwidth,trim={1cm 1cm 0.8cm 1cm},clip]{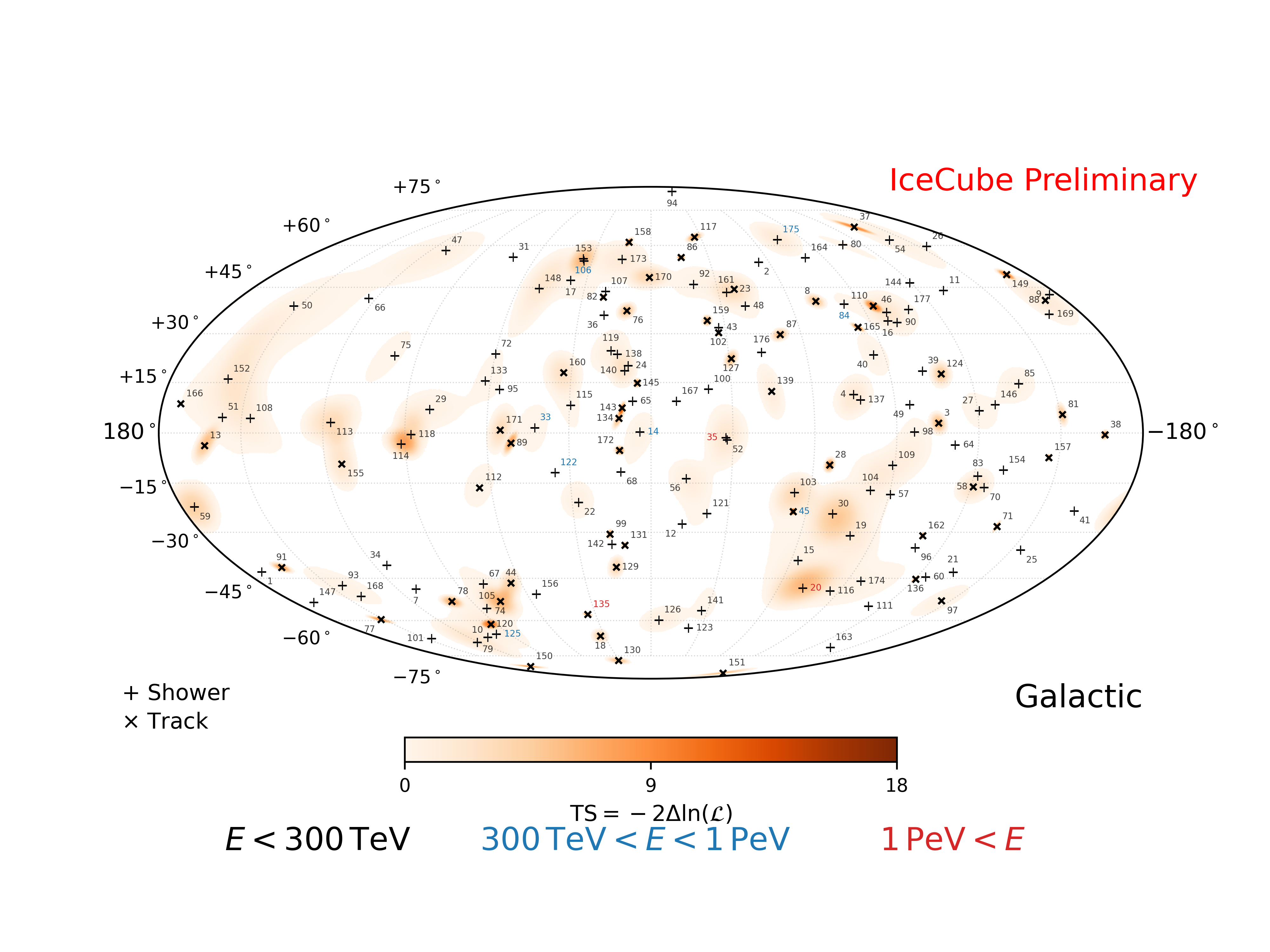}
\caption{The TS as evaluated at points across a HEALPix map with nside=512~\cite{Gorski:2004by} is indicated by the color scale. Events that were reconstructed with the shower (track) algorithm are indicated with $+$ ($\times$). They are also indicated by their event IDs in the skymap, which are color coded according to their reconstructed visible energy.}
\label{fig:skymap}
\end{figure*}
The all sky TS map is shown in Fig.~\ref{fig:skymap} in galactic coordinates. Due to the limited statistics of the sample, we do not expect to observe significant emission from point sources after trials correction. After accounting for trials, a p-value of $p=0.17$ is obtained, consistent with existing results.

\begin{figure*}[hbt]
\centering
\includegraphics[width=0.8\textwidth]{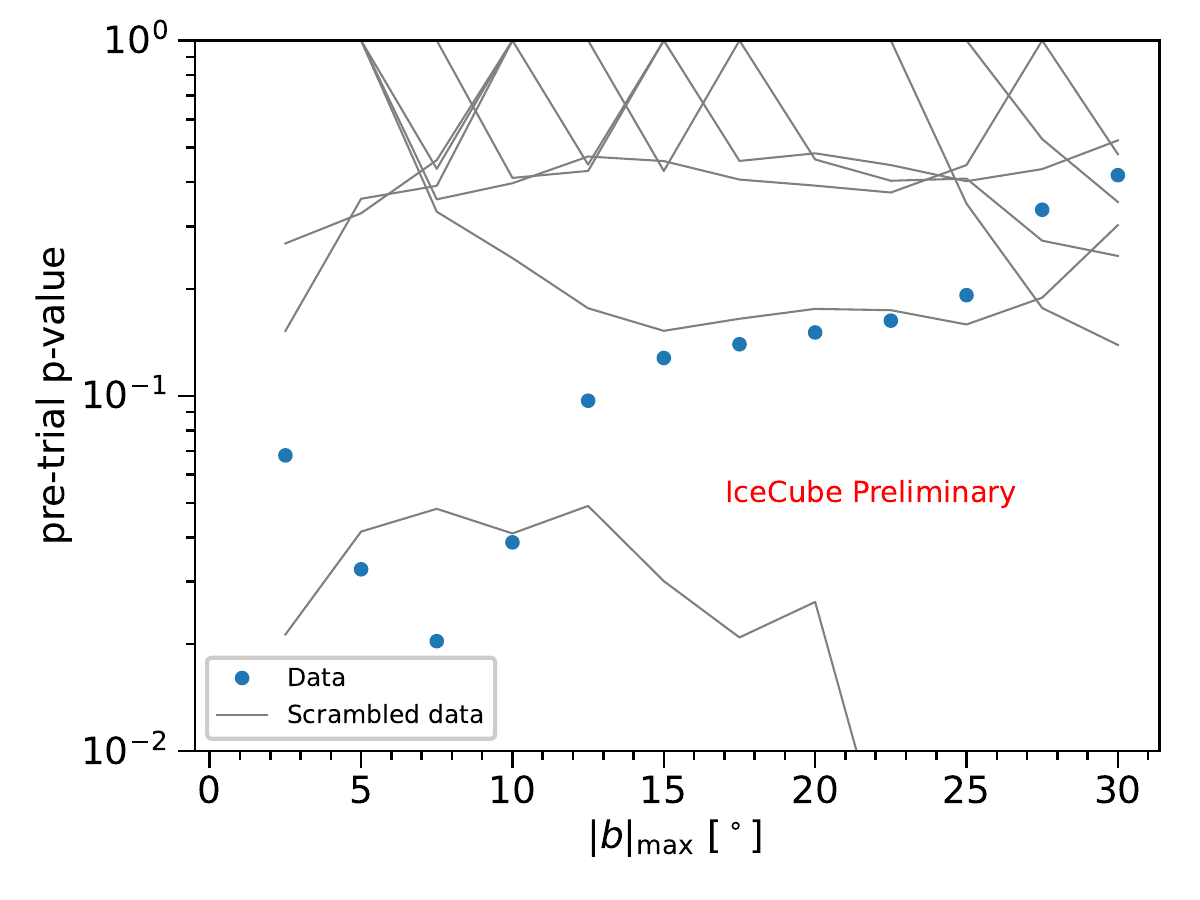}
\caption{Pre-trial p-values obtained for data (blue points) at the $|b|_{\mathrm{max}}$ values indicated describing the width of a box template across the galactic plane. The same is shown for 10 pseudoexperiments under the background-only hypothesis (gray lines).}
\label{fig:pbox}
\end{figure*}
In addition to the all-sky analysis, two galactic plane clustering tests were performed. The first searched for galactic clustering within a band across the galactic plane, $|b|_{\mathrm{max}}$, where $b$ denotes the galactic latitude. The result is shown in Fig.~\ref{fig:pbox} (blue points) for the different widths that were tested, with the most significant p-value at $|b|_{\mathrm{max}}=7.5^\circ$ at $p=0.02 (0.067)$ pre-trial (post-trial) with $\hat{n}_s=13.4$. The grey lines show the pre-trial p-values obtained for 10 pseudoexperiments under the background-only hypothesis.

The second galactic plane clustering test assumed neutrino flux expectations from the Fermi-$\pi^0$ model. We find a significance of $p=0.065$ with $\hat{n}_s = 13.8$. Based on the sensitivity studies discussed in Sec.~\ref{sec:sensitivity}, this is consistent with the measurement reported in~\cite{doi:10.1126/science.adc9818} where the galactic contribution to the total diffuse flux is on the order of 10 percent. Based on the measured significance, an upper limit on the galactic contribution is obtained at the 90\% confidence level to be $f_{90}<27\%$ of the total diffuse neutrino flux~\cite{IceCube:2020wum}.

\section{Conclusion}\label{sec:outro}

An update to the IceCube HESE dataset is provided corresponding to 12 years of data-taking. An additional 62 events were selected, on top of the existing 102 events previously reported~\cite{IceCube:2020wum}. Recent updates to ice modeling~\cite{tc-2022-174,Chirkin:2023icrc} are taken into account by reconstructing all events with DirectFit, which fully simulates Cherenkov photon propagation through the ice~\cite{Chirkin:2013avz}. For each event, the appropriate reconstruction routine (track or shower) is predetermined based on the likelihood difference under the two hypotheses. With changes in the ice model and some reclassification, shifts in direction can be expected for some events when compared to~\cite{IceCube:2020wum} (Appendix J). The full, updated dataset can be accessed at~\cite{DVN/PZNO2T_2023}.

The highest energy event was reconstructed as a track to have a visible energy of \SI{4.8}{\peta \eV}. Assuming the track is due to a high-energy muon, and taking into account energy losses along the track while assuming a uniform sampling of initial muon energies~\cite{Chirkin:2004hz}, a preliminary estimate for the total energy of this event is \SI{13\pm5}{\peta \eV}.

Spatial clustering tests performed include an all sky scan and two tests for emission from the galactic plane. The most significant result after trials correction is obtained for the Fermi-$\pi^0$ template at a p-value of 6.5\%. Using a data-driven approach, this corresponds to a 90\% upper limit of $f_{90}<27\%$ galactic contribution to the total astrophysical neutrino flux given in~\cite{IceCube:2020wum}.

% Bibtex references:
\bibliographystyle{ICRC}
\bibliography{main}

\providecommand{\href}[2]{#2}\begingroup\raggedright\begin{thebibliography}{10}

\bibitem{IceCube:2013cdw}
{\bfseries IceCube} Collaboration, M.~G. Aartsen {\em et~al.}
  \href{http://dx.doi.org/10.1103/PhysRevLett.111.021103}{{\em Phys. Rev.
  Lett.} {\bfseries 111} (2013) 021103}.

\bibitem{IceCube:2013low}
{\bfseries IceCube} Collaboration, M.~G. Aartsen {\em et~al.}
  \href{http://dx.doi.org/10.1126/science.1242856}{{\em Science} {\bfseries
  342} (2013) 1242856}.

\bibitem{Kopper:2015vzf}
{\bfseries IceCube} Collaboration, C.~Kopper, W.~Giang, and N.~Kurahashi
  \href{http://dx.doi.org/10.22323/1.236.1081}{{\em PoS} {\bfseries ICRC2015}
  (2016) 1081}.

\bibitem{Kopper:2017zzm}
{\bfseries IceCube} Collaboration, C.~Kopper
  \href{http://dx.doi.org/10.22323/1.301.0981}{{\em PoS} {\bfseries ICRC2017}
  (2018) 981}.

\bibitem{IceCube:2020wum}
{\bfseries IceCube} Collaboration, R.~Abbasi {\em et~al.}
  \href{http://dx.doi.org/10.1103/PhysRevD.104.022002}{{\em Phys. Rev. D}
  {\bfseries 104} (2021) 022002}.

\bibitem{Chirkin:2013avz}
{\bfseries IceCube} Collaboration, D.~Chirkin, ``{Event reconstruction in
  IceCube based on direct event re-simulation},'' in {\em {33rd International
  Cosmic Ray Conference}}, p.~0581.
\newblock 2013.

\bibitem{tc-2022-174}
{\bfseries IceCube} Collaboration, R.~Abbasi {\em et~al.}
  \href{http://dx.doi.org/10.5194/tc-2022-174}{{\em The Cryosphere Discussions}
  {\bfseries 2022} (2022) 1--48}.

\bibitem{Chirkin:2023icrc}
{\bfseries IceCube} Collaboration, D.~Chirkin and M.~Rongen {\em PoS}
  {\bfseries ICRC2023} (these proceedings) 975.

\bibitem{kent1982}
J.~T. Kent {\em J. Roy. Statist. Soc. Ser. B} {\bfseries 44} no.~1, (1982)
  71--80.

\bibitem{Yuan:2020}
T.~Yuan \href{http://dx.doi.org/10.1007/s00180-020-01023-w}{{\em Comput. Stat.}
  {\bfseries 36} (Mar, 2021) 409--420}.

\bibitem{DVN/PZNO2T_2023}
{\bfseries IceCube} Collaboration, ``{IceCube HESE 12-year data release},''
  2023.
\newblock \url{https://doi.org/10.7910/DVN/PZNO2T}.

\bibitem{Basu:2023icrc}
{\bfseries IceCube} Collaboration, V.~Basu and A.~Balagopal~V. {\em PoS}
  {\bfseries ICRC2023} (these proceedings) 1007.

\bibitem{Silva:2023icrc}
{\bfseries IceCube} Collaboration, M.~Silva {\em PoS} {\bfseries ICRC2023}
  (these proceedings) 1008.

\bibitem{Naab:2023icrc}
{\bfseries IceCube} Collaboration, R.~Naab, E.~Ganster, and Z.~Zhang {\em PoS}
  {\bfseries ICRC2023} (these proceedings) 1064.

\bibitem{Aartsen:2013vja}
{\bfseries IceCube} Collaboration, M.~G. Aartsen {\em et~al.}
  \href{http://dx.doi.org/10.1088/1748-0221/9/03/P03009}{{\em JINST} {\bfseries
  9} (2014) P03009}.

\bibitem{YUAN2023168440}
{\bfseries IceCube} Collaboration, T.~Yuan
  \href{http://dx.doi.org/https://doi.org/10.1016/j.nima.2023.168440}{{\em
  Nucl. Instrum. Meth. A} (2023) 168440}.

\bibitem{Yuan:2023icrc}
{\bfseries IceCube} Collaboration, T.~Yuan {\em PoS} {\bfseries ICRC2023}
  (these proceedings) 1005.

\bibitem{Chirkin:2004hz}
D.~Chirkin and W.~Rhode, ``{Muon Monte Carlo: A High-precision tool for muon
  propagation through matter},'' 7, 2004.

\bibitem{Braun:2008bg}
J.~Braun, J.~Dumm, F.~De~Palma, C.~Finley, A.~Karle, and T.~Montaruli
  \href{http://dx.doi.org/10.1016/j.astropartphys.2008.02.007}{{\em Astropart.
  Phys.} {\bfseries 29} (2008) 299--305}.

\bibitem{Ackermann_2012}
M.~Ackermann {\em et~al.}
  \href{http://dx.doi.org/10.1088/0004-637X/750/1/3}{{\em The Astrophysical
  Journal} {\bfseries 750} no.~1, (Apr, 2012) 3}.

\bibitem{Gaggero_2015}
D.~Gaggero, D.~Grasso, A.~Marinelli, A.~Urbano, and M.~Valli
  \href{http://dx.doi.org/10.1088/2041-8205/815/2/L25}{{\em The Astrophysical
  Journal Letters} {\bfseries 815} no.~2, (Dec, 2015) L25}.

\bibitem{Gorski:2004by}
K.~M. G\'orski, E.~Hivon, A.~J. Banday, B.~D. Wandelt, F.~K. Hansen,
  M.~Reinecke, and M.~Bartelman \href{http://dx.doi.org/10.1086/427976}{{\em
  Astrophys. J.} {\bfseries 622} (2005) 759--771}.

\bibitem{doi:10.1126/science.adc9818}
{\bfseries IceCube} Collaboration, R.~Abbasi {\em et~al.}
  \href{http://dx.doi.org/10.1126/science.adc9818}{{\em Science} {\bfseries
  380} no.~6652, (2023) 1338--1343}.

\end{thebibliography}\endgroup

% Alternatively, you can include references by hand:
%\begin{thebibliography}{99}
%\bibitem{...}
%
%\end{thebibliography}

\clearpage

%The following list of authors, affiliations and funding agencies will be updated at the day of submission. The following template is a placeholder generated via https://authorlist.icecube.wisc.edu/icecube on March 25, 2023 and will be updated.
\section*{Full Author List: IceCube Collaboration}

\scriptsize
\noindent
R. Abbasi$^{17}$,
M. Ackermann$^{63}$,
J. Adams$^{18}$,
S. K. Agarwalla$^{40,\: 64}$,
J. A. Aguilar$^{12}$,
M. Ahlers$^{22}$,
J.M. Alameddine$^{23}$,
N. M. Amin$^{44}$,
K. Andeen$^{42}$,
G. Anton$^{26}$
% ...
\\
\\
% ...
$^{12}$ Universit{\'e} Libre de Bruxelles, Science Faculty CP230, B-1050 Brussels, Belgium \\
% ...
$^{17}$ Department of Physics, Loyola University Chicago, Chicago, IL 60660, USA \\
$^{18}$ Dept. of Physics and Astronomy, University of Canterbury, Private Bag 4800, Christchurch, New Zealand \\
% ...
$^{22}$ Niels Bohr Institute, University of Copenhagen, DK-2100 Copenhagen, Denmark \\
$^{23}$ Dept. of Physics, TU Dortmund University, D-44221 Dortmund, Germany \\
% ...
$^{26}$ Erlangen Centre for Astroparticle Physics, Friedrich-Alexander-Universit{\"a}t Erlangen-N{\"u}rnberg, D-91058 Erlangen, Germany \\
% ...
$^{40}$ Dept. of Physics and Wisconsin IceCube Particle Astrophysics Center, University of Wisconsin{\textendash}Madison, Madison, WI 53706, USA \\
% ...
$^{42}$ Department of Physics, Marquette University, Milwaukee, WI, 53201, USA \\
% ...
$^{44}$ Bartol Research Institute and Dept. of Physics and Astronomy, University of Delaware, Newark, DE 19716, USA \\
% ...
$^{63}$ Deutsches Elektronen-Synchrotron DESY, Platanenallee 6, 15738 Zeuthen, Germany  \\
$^{64}$ Institute of Physics, Sachivalaya Marg, Sainik School Post, Bhubaneswar 751005, India \\
% ...

\subsection*{Acknowledgements}

\noindent
The authors gratefully acknowledge the support from the following agencies and institutions:
USA {\textendash} U.S. National Science Foundation-Office of Polar Programs,
U.S. National Science Foundation-Physics Division,
U.S. National Science Foundation-EPSCoR,
Wisconsin Alumni Research Foundation,
Center for High Throughput Computing (CHTC) at the University of Wisconsin{\textendash}Madison,
Open Science Grid (OSG),
Advanced Cyberinfrastructure Coordination Ecosystem: Services {\&} Support (ACCESS),
Frontera computing project at the Texas Advanced Computing Center,
U.S. Department of Energy-National Energy Research Scientific Computing Center,
Particle astrophysics research computing center at the University of Maryland,
Institute for Cyber-Enabled Research at Michigan State University,
and Astroparticle physics computational facility at Marquette University;
Belgium {\textendash} Funds for Scientific Research (FRS-FNRS and FWO),
FWO Odysseus and Big Science programmes,
and Belgian Federal Science Policy Office (Belspo);
Germany {\textendash} Bundesministerium f{\"u}r Bildung und Forschung (BMBF),
Deutsche Forschungsgemeinschaft (DFG),
Helmholtz Alliance for Astroparticle Physics (HAP),
Initiative and Networking Fund of the Helmholtz Association,
Deutsches Elektronen Synchrotron (DESY),
and High Performance Computing cluster of the RWTH Aachen;
Sweden {\textendash} Swedish Research Council,
Swedish Polar Research Secretariat,
Swedish National Infrastructure for Computing (SNIC),
and Knut and Alice Wallenberg Foundation;
European Union {\textendash} EGI Advanced Computing for research;
Australia {\textendash} Australian Research Council;
Canada {\textendash} Natural Sciences and Engineering Research Council of Canada,
Calcul Qu{\'e}bec, Compute Ontario, Canada Foundation for Innovation, WestGrid, and Compute Canada;
Denmark {\textendash} Villum Fonden, Carlsberg Foundation, and European Commission;
New Zealand {\textendash} Marsden Fund;
Japan {\textendash} Japan Society for Promotion of Science (JSPS)
and Institute for Global Prominent Research (IGPR) of Chiba University;
Korea {\textendash} National Research Foundation of Korea (NRF);
Switzerland {\textendash} Swiss National Science Foundation (SNSF);
United Kingdom {\textendash} Department of Physics, University of Oxford.

\end{document}